# Regulation of Signal Duration and the Statistical Dynamics of Kinase Activation by Scaffold Proteins

Jason W. Locasale[1], Arup K. Chakraborty[1,2,3]*

1 Department of Biological Engineering, Massachusetts Institute of Technology, Cambridge, Massachusetts, United States of America, 2 Department of Chemical Engineering, Massachusetts Institute of Technology, Cambridge, Massachusetts, United States of America, 3 Department of Chemistry, Massachusetts Institute of Technology, Cambridge, Massachusetts, United States of America

## Abstract

Scaffolding proteins that direct the assembly of multiple kinases into a spatially localized signaling complex are often essential for the maintenance of an appropriate biological response. Although scaffolds are widely believed to have dramatic effects on the dynamics of signal propagation, the mechanisms that underlie these consequences are not well understood. Here, Monte Carlo simulations of a model kinase cascade are used to investigate how the temporal characteristics of signaling cascades can be influenced by the presence of scaffold proteins. Specifically, we examine the effects of spatially localizing kinase components on a scaffold on signaling dynamics. The simulations indicate that a major effect that scaffolds exert on the dynamics of cell signaling is to control how the activation of protein kinases is distributed over time. Scaffolds can influence the timing of kinase activation by allowing for kinases to become activated over a broad range of times, thus allowing for signaling at both early and late times. Scaffold concentrations that result in optimal signal amplitude also result in the broadest distributions of times over which kinases are activated. These calculations provide insights into one mechanism that describes how the duration of a signal can potentially be regulated in a scaffold mediated protein kinase cascade. Our results illustrate another complexity in the broad array of control properties that emerge from the physical effects of spatially localizing components of kinase cascades on scaffold proteins.

Funding: This work was funded through an NIH Director's Pioneer Award and NIH grant PO1 AI071195-01 to AKC.

Competing Interests: The authors have declared that no competing interests exist.

* E-mail: arupc@MIT.edu

## Introduction

In the context of signal transduction, cells integrate signals derived from membrane proximal events and convert them into the appropriate cell decision. Within the complex networks that integrate these signals lies a highly conserved motif involving the sequential activation of multiple protein kinases. Signal propagation through these kinase cascades is often guided by a scaffolding protein that assembles protein kinases into a multi-protein complex. Signaling complexes maintained by scaffolds are intensely studied and have been shown to affect myriad cell decisions [1–7]. Despite numerous advances in the understanding of the signaling function of scaffold proteins [8–15], many questions remain. For instance, although scaffolds are believed to have profound effects on the dynamics of signal propagation [6,9,10,16], the mechanisms that underlie how scaffolds regulate signaling dynamics are not well understood.

One key factor in specifying a cellular decision is the duration of a signal (i.e. the time over which a kinase remains active) [17,18]. Differences in signal duration have been implicated as the basis of differential decisions in myriad cell processes. For example, it has been suggested that decisions on growth factor induced cell proliferation, positive and negative selection of T cells, apoptotic programs, cell cycle progression, among many others, are regulated by the duration of signaling [19–24]. Therefore, the issue of how a signal output, such as the activity of extracellular regulatory kinase (ERK) in a MAPK pathway, is distributed over time, is of considerable interest.

There are many ways in which the duration of the output of a kinase cascade can be controlled. Regulation of signaling dynamics can arise from processes upstream of the cascade [25]. For example, degradation of upstream signaling components such as the surface receptors [26] and differential kinetics of GTPase regulators [27,28] can be essential in regulating MAPK signaling dynamics [25]. Also, multisite phosphorylation is predicted to influence signal duration [29]. It has been also been shown that differential modes of feedback regulation that are manifested under different conditions within the same cascade can regulate signal duration [30]. Scaffold proteins have also been implicated as key determinants in the regulation of signal duration [9,10,31].

Because the many factors that control scaffold mediated signaling are difficult to systematically control in a laboratory setting, a precise understanding of how scaffold proteins affect the dynamics of signal transduction has proven elusive. Computational models have been useful in understanding some of the many complex ways in which scaffolds influence signal transduction [16,32–34]. However, it is currently impossible to model theoretically all aspects of any biological signaling process—computational models ultimately require that many gross simplifications be made. Our aim is, therefore, not to attempt to simulate every detail of a specific biochemical pathway but rather investigate the consequences that emerge from a simple scenario of

**Author Summary**

Signal transduction is the science of cellular communication. Cells detect signals from their environment and use them to make decisions such as whether or when to proliferate. Tight regulation of signal transduction is required for all healthy cells, and aberrant signaling leads to countless diseases such as cancer and diabetes. For example, in higher organisms such as mammals, signal transduction that leads to cell proliferation is often guided by a scaffold protein. Scaffolding proteins direct the assembly of multiple proteins involved in cell signaling by providing a platform for these proteins to carry out efficient signal transmission. Although scaffolds are widely believed to have dramatic effects on how signal transduction is carried out, the mechanisms that underlie these consequences are not well understood. Therefore, we used a computational approach that simulates the behavior of a model signal transduction module comprising a set of proteins in the presence of a scaffold. The simulations reveal mechanisms for how scaffolds can dynamically regulate the timing of cell signaling. Scaffolds allow for controlled levels of signal that are delivered inside the cell at appropriate times. Our findings support the possibility that these signaling dynamics regulated by scaffolds affect cell decision-making in many medically important intracellular processes.

scaffold mediated signaling whereby a model cascade assembles onto a scaffold. In modeling this scenario in itself, we hope to learn more about the functional and mechanistic consequences that these specific physical constraints, imposed by assembling components of a biochemical cascade onto a scaffold, confer to signaling pathways. In parsing these effects from the myriad others that are undoubtedly important, our hope is that our results can serve as a framework for understanding the extent to which these effects are important in specific biological contexts such as the Mitogen Activated Protein Kinase (MAPK) pathway.

One theoretical analysis of scaffold mediated cell signaling revealed the presence of non mononotic behavior in signal output as a function of scaffold concentration [34]. If scaffolds are required for signaling, then too few scaffolds will be detrimental to signaling. On the other hand, if scaffolds are present in excess, signaling complexes become incompletely assembled and the signal output is attenuated. As a consequence of this "prozone" effect, scaffolds were shown to also differentially affect the kinetics of signaling.

The observation that scaffolds can differentially affect signaling dynamics leads to many questions. How do scaffold proteins control the time scales involved in signal propagation? An important metric of cell signaling is the time it takes for a downstream kinase to become active [35,36]. As signal transduction is stochastic in nature, the more precise question is: what is the distribution of times characterizing the activation of a downstream kinase? How do scaffolds affect this distribution, and what might be the biological consequences of changes in this distribution as a result of signaling on a scaffold? We compute first passage time distributions [37] using a stochastic computer simulation method to investigate these questions.

Specifically, we use a kinetic Monte Carlo algorithm. We have previously used such methods to study a different question concerning the regulation of signal amplitude by scaffold proteins [33]. It is also possible that a differential equation model that considers mean-field kinetics could be used to study the first passage time distribution [37]. However, such an approach would require the imposition of absorbing boundary conditions that can make the numerical analysis difficult.

Our simulation results suggest that, depending on physiological conditions, scaffold proteins can allow kinase cascades to operate in different dynamical regimes that allow for large increases and decreases in the speed and characteristic time scale of signal propagation. Furthermore, and perhaps more importantly, scaffolds are shown to influence the statistical properties of the times at which kinases are activated in complex ways. Scaffolding protein kinases cascades can allow for broadly distributed waiting times of kinase activation, whereas in the absence of a scaffold, the time it takes for a kinase to be activated is effectively characterized by a single time scale. These stochastic characteristics of scaffold-mediated kinase cascades are, to our knowledge, elucidated for the first time and may have diverse biological consequences that pertain to how signal duration is regulated. It is also our hope that our results provide a framework for achieving a deeper qualitative understanding of how scaffolding proteins can regulate the dynamics of cell signaling and the statistical properties of signal transduction.

## Results

### Model of a Protein Kinase Cascade

For our study, we considered a model three tiered protein kinase cascade such as the MAPK pathway [38]. Since our aim is to study the effects of spatially localizing protein kinases on signaling dynamics, we considered a minimal description of a model kinase cascade. Many factors that are undoubtedly important in regulating signaling dynamics were not considered. These factors include feedback regulation within the cascade, allosteric and or catalytic functions provided by the scaffold, and the effects of multiple phosphorylations of each kinase [11,25,26,30,39,40].

In our model, signal propagation occurs in a three step hierarchical fashion: an initial stimulus (S) activates a MAP3K (A) that in turn, activates a MAP2K (B), that subsequently can activate its MAPK (C) substrate—phosphatases can deactivate each activated species and this deactivation occurs regardless of whether or not the active kinase is bound to a scaffold. A schematic is presented in Figure 1A that illustrates the basic processes that are allowed in our model. A steady-state ensemble is considered. That is, simulations are allowed to first reach a dynamic steady-state and once this state is reached, dynamics are studied. We do not consider dynamics from the starting time that requires propagation through a hierarchical cascade.

Recent work has studied the statistical dynamics of kinase activation that result from the hierarchical organization of a kinase cascade; in that study, it was shown that the hierarchical structure of the cascade gives rise to broad waiting time distributions of cascade activation. In the regime that we study here, these effects are absent since activation of the cascade requires that an inactive C protein encounter an active B protein; our motivation is thus to investigate how the dynamics of kinase activation can be affected by assembling components of the cascade onto a scaffolding protein that localizes single complexes. Therefore, we do not emphasize how the hierarchical structure of a signaling cascade effects signal propagation and instead focus on how assembly of the cascade onto a scaffold affects signaling dynamics. We also underscore the notion that in our approach, many undoubtedly important effects such as the hierarchical structure of protein kinase cascades, the influence of feedback loops, differential enzymatic mechanisms and allosteric control by scaffolds are neglected. Again, by excising these effects, we restrict our attention to a hypothetical scenario that aims only to investigate the consequences of assembling components of a cascade onto a scaffold protein.

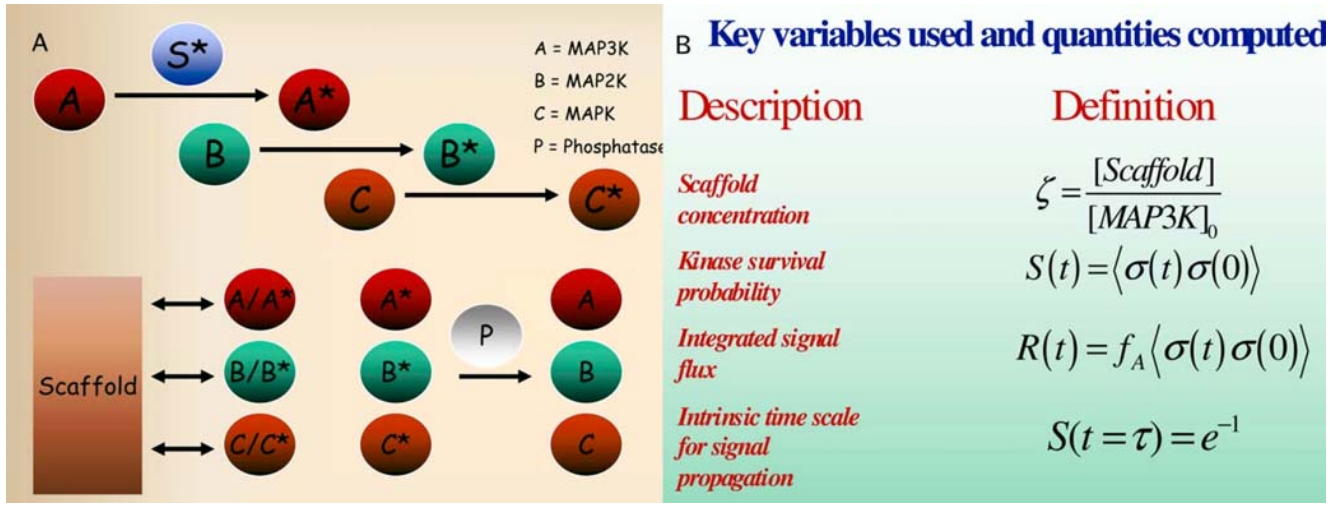

**Figure 1. A model to study dynamical properties of a scaffold mediated signaling cascade.** (A) Schematic of the events considered in the scaffold mediated signaling cascade. Each kinase, if activated, can activate its downstream substrate when the two proteins are in close proximity. Kinases can bind and unbind to the scaffold and phosphatases can, upon encountering an active kinase, deactivate it. Activation potentially occurs both in solution and on a scaffold. Each forward and backward reaction is modeled as a elementary reactive collision with an energy barrier, E. Energy barriers, E, were taken to be zero so that all kinetics are diffusion limited (B) Key variables and the main quantities computed are shown. $\zeta$ is the dimensionless scaffold concentration. The concentration of scaffold proteins [Scaffold] is scaled to the density of the first kinase in the cascade $[MAP3K]_0$ ($\zeta \equiv \frac{[Scaffold]}{[MAP3K]_0}$). The survival probability $S(t) \equiv \langle \sigma(t)\sigma(0) \rangle$, where $\sigma(t)$ is zero is a kinase has become active and one otherwise (brackets denote an ensemble average); $S(t)$ is the probability that the final kinase has been activated given that it was inactive at $t=0$. $R(t) \equiv f_A S(t) = f_A \langle \sigma(t)\sigma(0) \rangle$, where $f_A$ is the fraction of active kinases at steady state; $R(t)$ is the integrated reactive flux of kinase activation. The characteristic time scale of signal propagation $\tau$ is defined by the relation $S(t=\tau) = e^{-1}$.
doi:10.1371/journal.pcbi.1000099.g001

The key quantities computed and parameters used are discussed below in Table 1 and Figure 1B. Additional details are provided in the Methods section.

**Table 1.** Notation and parameters used.

| Index $i$ | Species Type | Index $j$ | Chemical State |
|---|---|---|---|
| $i=0$ | Stimulus (S) | $j=1$ | Unbound, inactive |
| $i=1$ | MAPKKK (A) | $j=2$ | Unbound, active |
| $i=2$ | MAPKK (B) | $j=3$ | Bound, inactive |
| $i=3$ | MAPK (C) | $j=4$ | Bound, active |
| $i=4$ | Phosphatase (P) | | |
| **Parameter Used** | **Description** | **Value** | |
| | Energy barrier for: | | |
| $E_{1,3}$, $E_{2,4}$ | Binding | $0 k_B T$ | |
| $E_{3,1}$, $E_{4,2}$ | Unbinding | $12 k_B T$ | |
| $E_{1,2}$, $E_{3,4}$ | Catalytic activation | $0 k_B T$ | |
| $E_{2,1}$, $E_{4,3}$ | Catalytic deactivation | $12 k_B T$ | |
| $(k^{eff})^{-1}$ | Reaction time scale | 1 | |
| $(D^{eff})^{-1}$ | Diffusion time scale | 10 | |

*All other values of $E_{j,j'}$ are taken to be infinite.
doi:10.1371/journal.pcbi.1000099.t001

## The Concentration of Scaffold Proteins Sets Time Scales for Signal Propagation Through a Kinase Cascade

To set the context, consider the consequences of signaling in two limiting cases in our model. When the binding affinity of the kinases to the scaffold, E, is low (defined here to be close to the thermal energy, $E \sim k_B T$; $k_B$ is Boltzman's constant and T is the temperature) and kinases disassociate rapidly from the scaffold, few proteins on average are bound to a scaffold. Therefore, signaling dynamics corresponds to that of a kinase cascade in solution. For a very strong affinity, $E \gg k_B T$ all available binding sites to scaffold proteins are occupied by kinases (on average). In this case, signaling dynamics are controlled by the time required for initial stimuli to encounter and interact with each fully assembled complex.

Therefore, we consider cases in which kinases can disassociate from their scaffolds and exchange with unbound kinases on time scales pertinent to cell signaling processes. Such time scales correspond to disassociation constants ($K_d$) on the order of 1–10 μM and off rates, $k_{off} \sim 1 s^{-1}$. Such $K_d$ values correspond to free energies of binding of roughly 7–9 kcal/mol, an energy scale typical of protein-protein interactions in kinase cascades [41]. We have used 12 $k_B T$ as the binding energy in our simulations which corresponds to ~7.2 kcal/mol. We also discuss the robustness of our results with respect to changes in this value. Scaffold concentration has been identified as a key variable that can regulate the efficiency of signal propagation through a kinase cascade [2,5,34]. For the set of parameters used in the simulations (Table 1), signal output (defined as the average steady state value of the final kinase in the cascade) has a non-monotonic (biphasic) dependence on the relative concentration of scaffolds $\zeta$ ($\left(\zeta \equiv \frac{[Scaffold]}{[MAP3K]_0}\right)$, where [Scaffold] is the concentration of the scaffold and $[MAP3K]_0$ is the concentration of the first kinase in the cascade) and peaks at an optimal value of $\zeta = 1$ [33,34].

To quantify signaling dynamics, we consider a survival probability $S(t)$ (methods) that, as mentioned, can be viewed as a type of autocorrelation function.

$$S(t) \sim \langle \sigma(t)\sigma(0) \rangle,$$

where $\sigma(t)$ equals 0 or 1 depending upon the activity of the final kinase within the cascade (methods) and the brackets indicate an average over all kinases in the simulation averaged over many simulations. This quantity gives the probability that the final kinase in the cascade remains inactive at time t given that it was inactive at time $t = 0$. Therefore, signaling dynamics can be monitored by observing the decay of this function with time.

In Figure 2A, $S(t)$ is computed for different values of the relative scaffold concentration, $\zeta$. The intrinsic time of signal propagation, $\tau$, is the value at which $S(t)$ decays to $e^{-1}$ of its original value ($S(t=\tau) = e^{-1}$). Upon increasing scaffold concentration, $\tau$ increases. At very high scaffold expression levels, signals propagate so slowly that cell signaling is not observed on experimentally measurable time scales which we take to be in our simulations $\gg 10^6$ Monte Carlo (MC) steps; 1 MC step$\sim$1 µs assuming a lattice spacing of 10 nm and a diffusion coefficient of 10 µm$^2$/s [42]. The increase in $\tau$ spans several orders of magnitude as is observed in Figure 2B. Distinct stages are also observed in the behavior of $\tau$, and are separated by an inflection point occurring shortly past the optimal value of scaffold concentration ($\zeta\sim1$). This phenomenon suggests that different physical processes are determining the signaling dynamics at different ranges of scaffold concentration.

These results also suggest that the concentration of scaffold proteins can in principle set an intrinsic time scale that determines the speed of signal propagation. Such an intrinsic time scale arises solely from changes in the concentration of scaffold proteins. This time scale can span several orders of magnitude for biologically relevant affinities and diffusion coefficients and increases monotonically with increasing scaffold concentration.

Note that these calculations consider only the speed of signaling and do not necessarily imply that signaling is more efficient when $\tau$ is small. To observe the total amount of integrated signal flux, the survival probability is conditioned with the probability that a kinase in the pool of signaling molecules is active in the steady state. We compute $R(t)$ defined as $S(t)$ multiplied by the average number of (the final downstream) kinases active at steady state,

$$R(t) \equiv f_A \langle \sigma(t)\sigma(0) \rangle,$$

where $f_A$ is the fraction of active kinases at steady state. The time derivative, $-\frac{d}{dt}R(t)$, can be thought of as a flux of activated kinases being produced. In Figure 2C, $R(t)$ is plotted as a function of time. For low concentrations of scaffolds, the small amount of signal, albeit quickly propagating, is rapidly quenched. As scaffold concentration increases, both the amplitude and duration of the signal increase up to an optimal value. Past the optimal value, higher scaffold concentrations result in signals with small amplitude but the duration of signaling is extended. The behavior of the integrated reactive flux is a direct consequence of the existence of an optimal scaffold concentration and "bell shaped" titration curve since the area under these curves is proportional to the average signal output [33,34].

## Scaffold Proteins Influence the Duration of Signaling by Controlling How Kinase Activation Is Distributed Over Time

Figure 2 emphasizes how the characteristic time for signal propagation is influenced by changes in the relative scaffold concentration. It also appears that the qualitative features of $S(t)$ change as scaffold concentration is varied. The decay of some distributions appears highly concentrated at a particular time while the decay of other distributions appears more broadly distributed.

To further investigate this observation, we plotted the survival probability as a function of the dimensionless time, $t/\tau$. If the decay of $S(t)$ is purely exponential, then $S(t/\tau)$ will have the form $e^{-t/\tau}$. Figure 3 shows $S(t/\tau)$ for different values of scaffold concentration and a decaying exponential function is given as a reference. One notices that $S(t/\tau)$ is exponential at negligible scaffold concentrations. As scaffold concentration increases, the behavior of $S(t/\tau)$ deviates from a single exponential decay. Near $\zeta = 1$, $S(t/\tau)$ shows maximal deviation from purely exponential kinetics. As scaffold expression increases past this point, the shape of $S(t/\tau)$ reverts back to an exponential form.

A deviation from exponential behavior can be quantified by considering a stretched exponential function,

$$e^{-(t/\tau)^\beta},$$

and fitting $S(t/\tau)$ to this form for different values of $\zeta$. One desirable feature of the stretched exponential function is the minimal number of parameters, $\tau$ and $\beta$, that are involved in the least-squares fit; also, the values of these parameters can be physically interpreted. $\tau$ gives the characteristic time for one overall timescale of signal propagation, and $\beta$ is a measure of how much the function, $S(t)$, deviates from a single exponential and thus how broadly distributed are the signaling dynamics. Figure 3B shows how $\beta$ depends on scaffold concentration. For these simulations, $\beta\sim1$ for small and large values of scaffold concentrations indicating exponential behavior. For intermediate values, $\beta$ peaks at a minimum of $\beta\sim0.6$, a significant deviation from purely exponential behavior.

In the limits of small and large scaffold concentrations, the presence of a single exponential decay, $\beta\sim1$ indicates that signal propagation, or the relaxation of $S(t/\tau)$, occurs at one characteristic time scale. In the intermediate regime, $\beta$ shows significant deviations from one, thus allowing for a broadly distributed signal. When $\beta$ is significantly less than one, signals can steadily propagate over several decades. In this regime, the waiting time distribution $f(t)$,

$$f(t) \equiv -\frac{dS(t)}{dt} = \frac{\beta}{t}\left(\frac{t}{\tau}\right)^\beta e^{-(t/\tau)^\beta},$$

has a large tail and the activation of kinases is slowly maintained over many time scales.

## A Multistate Kinetic Mechanism Illustrates the Competition Between the Many Time Scales Involved in Scaffold-Mediated Cell Signaling

Why do we observe exponential and non-exponential behavior under different conditions? Signal transduction in our model occurs on a time scale that is much slower than the microscopic time scales associated with diffusion, binding/unbinding, and enzyme catalysis. We might therefore expect that some coarse-graining exists whereby events at these fast, "microscopic" time scales interact with other relevant biophysical parameters (e.g. scaffold concentration) to give rise to emergent properties that evolve on slower times scales. These processes are a manifestation of the collective dynamics of the many processes that occur on faster time scales. Understanding the factors that govern these emergent time scales would then provide insight into the origin of the different temporal characteristics that are revealed by our simulations.

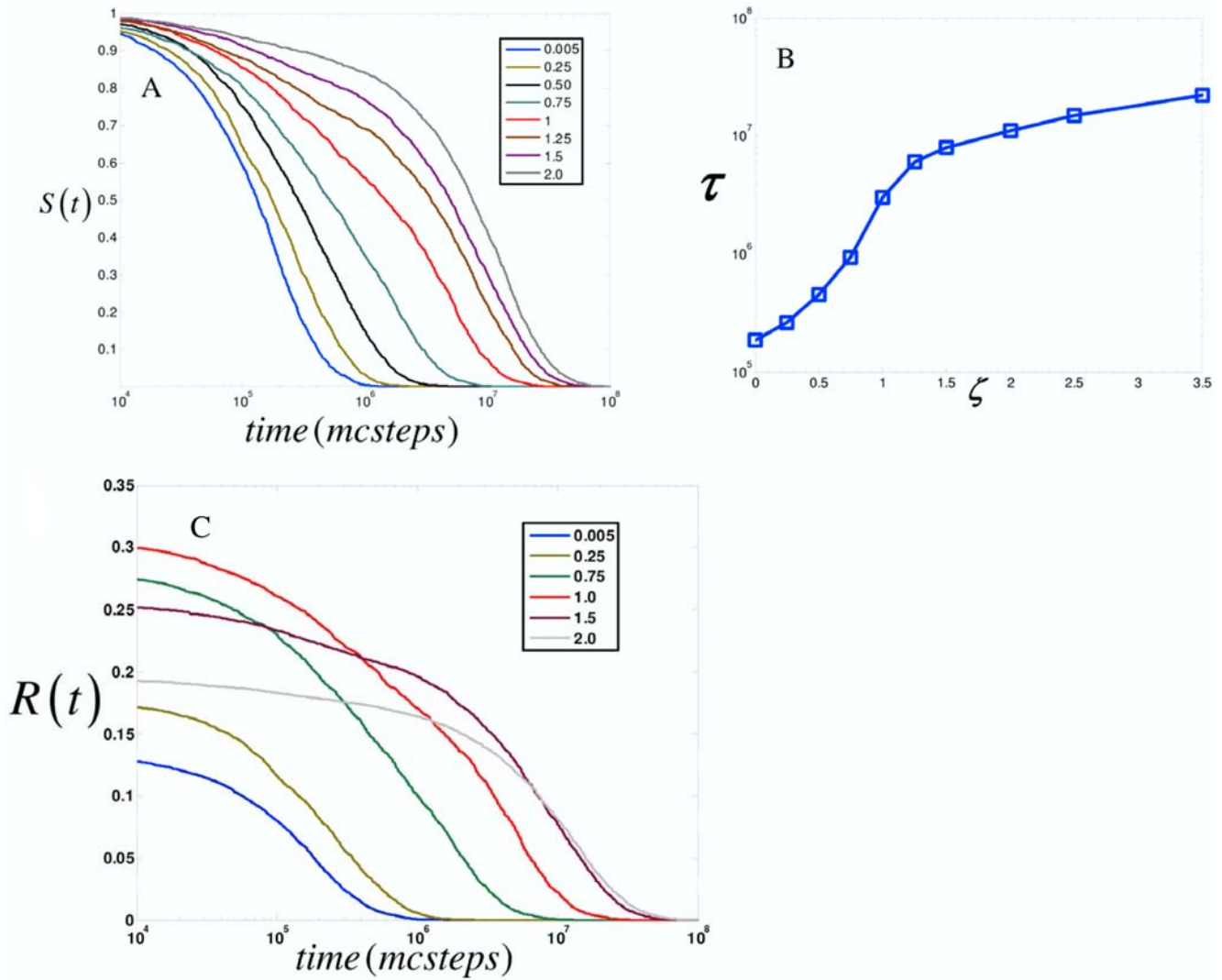

**Figure 2. The concentration of scaffold proteins sets time scales for signal propagation.** (A) $S(t)$ as a function of time for different values of relative scaffold concentration, $\zeta$ ($\zeta \equiv \frac{[Scaffold]}{[MAP3K]_0}$). $\zeta$ ranges from 0.005 to 2 times the optimal value. (B) The characteristic time scale $\tau$ ($S(t=\tau) = e^{-1}$) is extracted from the curves in (A), and its variation with $\zeta$ is shown. Two regimes are observed and are separated by an inflection point. (C) Integrated signal flux $R(t)$ for different values of $\zeta$.
doi:10.1371/journal.pcbi.1000099.g002

In order for a signal to propagate (i.e. for the last kinase in the cascade to become active), a hierarchical sequence of phosphorylation reactions among kinases must occur that leads to the final kinase in the cascade being activated by its upstream kinases. The activation process may occur either in solution or on a scaffold. Also, in the course of signaling, kinases can exchange from a scaffold. Some kinases are bound to a scaffold that contains an incomplete assembly of the necessary signaling molecules, and are not signaling competent. Ultimately, an inactive kinase can exist in one of three states: in solution, bound to a complete complex, or bound to an incomplete complex. Figure 4A contains a diagram of such a minimal picture and arrows denote transitions between the four states.

This minimalist description clarifies the behavior in Figures 3A and 3B. For low scaffold concentrations ($\zeta \ll 1$), kinases predominately exist in solution and signal transduction is dominated by the time it takes for an upstream kinase to encounter its downstream enzyme. Since a steady-state ensemble is used, the rate limiting step for signal propagation is the diffusion limited collision between an active B* molecule with an inactive C molecule. For high scaffold concentrations ($\zeta \gg 1$), kinases predominately exist in incomplete signaling complexes and signal transduction is limited by a time scale that characterizes the turnover of a signaling incompetent complex to one that is able to signal. For intermediate concentrations, inactive kinases can exist in each of three states and transitions between these states also occur. Thus, the source of the nonexponential relaxation (i.e. $\beta<1$) arises from the mixing of many time scales that are relevant for intermediate scaffold concentrations. Figure 4B illustrates this minimal picture of the kinetics of signal propagation derived from these physical considerations.

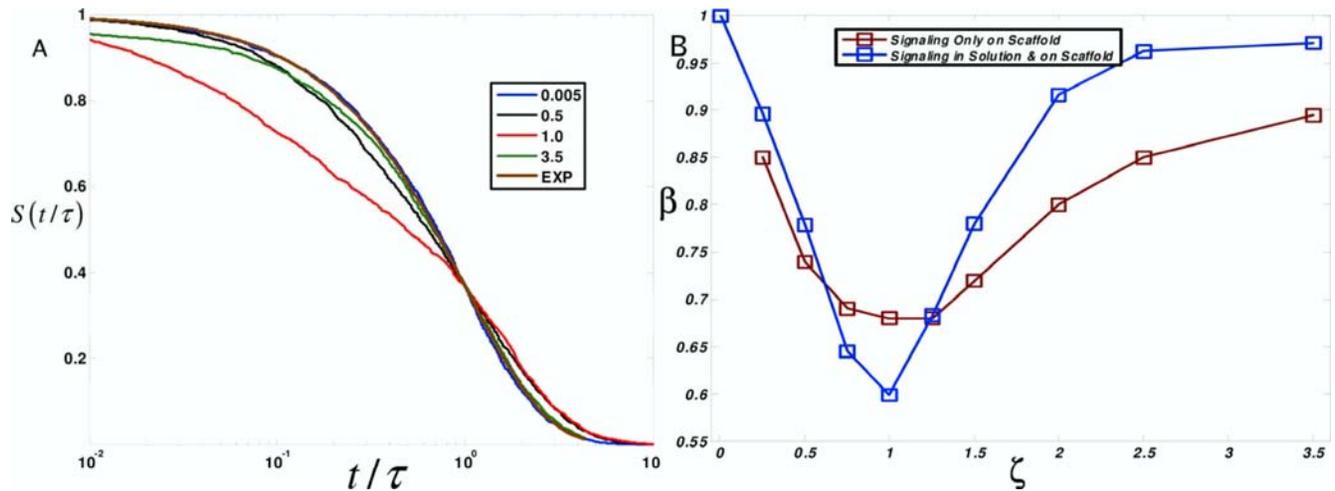

**Figure 3. Scaffold proteins allow for signals to be distributed over many time scales.** Variation of the survival probability with time scaled to characteristic time scales, $\tau$. (A) $S(t/\tau)$ (on a semi-logarithmic scale) for different values of $\zeta$. Values of $\zeta$ are given in the legend. Large deviations of exponential decay are observed near the optimal value of $\zeta$ ($\zeta = 1$, red). (B) Survival probabilities were fit to a stretched exponential function ($S(t/\tau) \sim e^{(t/\tau)^\beta}$). Values of stretching exponent $\beta$ as a function of scaffold density $\zeta$ are shown. Two cases are considered: (1) kinases can be activated only while bound to a scaffold (red) and (2) kinases can be activated while in solution and bound to a scaffold (blue). $\beta$ deviates most from a purely an exponential ($\beta = 1$) at the optimal value of scaffold density ($\zeta = 1$) in both cases.
doi:10.1371/journal.pcbi.1000099.g003

Also note that the sensitivity of our results to changes in model parameters can be understood from this simple picture of scaffold mediated signaling dynamics. For instance, changes in kinase and scaffold concentrations result in changes in the relative amount of kinases existing in the three states in ways that have been previously characterized [33,34]. Changes to other parameters such as the rates of activation and deactivation and the concentration of phosphatases alter the rates of transitions between these different states. For instance, if phosphatase concentrations are very large, then activation in solution is very slow and occurs predominantly on a scaffold. Also, slower rates of activation (and larger rates of deactivation) result in a larger portion of signaling originating from kinases that are bound to scaffolds. In general, when the activation of kinases originates more (less) predominantly from a particular state in the minimal model, $\beta$ increases (decreases). When multiple pathways to kinase activation contribute with comparable time scales, $\beta$ is small, and signaling is broadly distributed over many time scales. We have performed many simulations with varying parameters to test the robustness and parameter sensitivity of our findings and find that that the qualitative behavior of our results follow this simple, qualitative, physical picture.

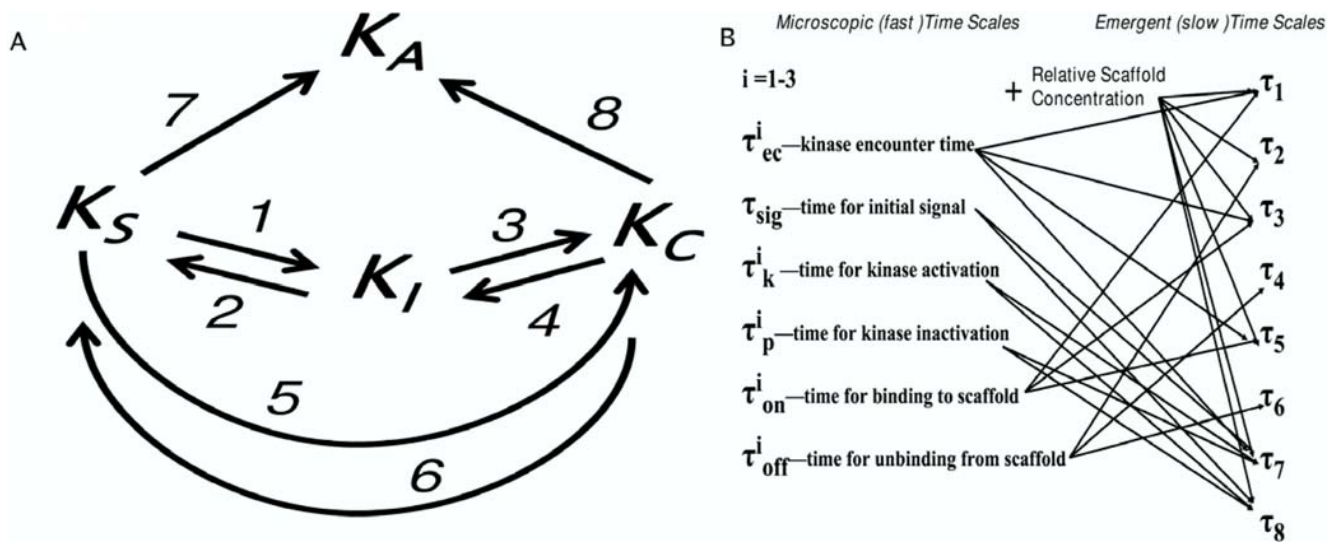

**Figure 4. Dynamics can be characterized by a multi-state kinetic mechanism.** Important time scales in scaffold mediated signaling. (A) Graph of multi-state kinetic model whose dynamics are governed by 8 transitions. Each kinase can transition between four states denoted with four subscripts: in solution (S), bound to a signaling competent complex (C), bound to a signaling incompetent complex (I), and activated (A). (B) Diagram depicting how the various processes occurring at fast time scales couple with scaffold concentration $\zeta$ to give rise to collective behavior occurring at slower time scales.
doi:10.1371/journal.pcbi.1000099.g004

Additional insight can be gleaned from consideration of the power spectrum of $S(t)$. The power spectrum,

$$P(\omega) \equiv |C(\omega)|^2$$

where,

$$C(\omega) = \int e^{i\omega t} S(t) dt$$

, computed in the frequency domain, resolves the time scale dependence of kinase activation. This approach has proven useful in studying the dynamics of complex biochemical networks in many contexts [43–45]. We first note that $S(t)$ obtained from the simulations fits well to the functional form $e^{-(t/\tau)^\beta}$ ($\chi^2$ values small). Thus, we use the parameters $\beta$ and $\tau$ that were extracted from the fits at low ($\zeta = 0.001$), optimal ($\zeta = 1.0$), and high ($\zeta = 3.5$) scaffold concentrations to compute $P(\omega)$ for these three cases.

In Figure 5, $(\tau^{opt})^{-2} P(\omega\tau^{opt})$ is plotted versus $\omega\tau^{opt}$ where the time $\tau^{opt}$ is the characteristic time scale $\tau$ for relaxation at the optimal $\zeta = 1$ scaffold concentration. That is, time is rescaled to units of $\tau^{opt}$. For each curve, at low $\omega\tau^{opt} \ll 1$ frequencies $P(\omega\tau^{opt})$ is constant $(P(\omega\tau \to 0) \to \tau^2)$ signifying that kinase activation has become uncorrelated. At high $\omega\tau^{opt} \gg 1$ frequencies, kinase activation is correlated and a power law decay is observed for each curve $P(\omega\tau^{opt}) \sim \omega^{-2}$. As a reference, note that for an exponential decay, $S(t) = e^{-t/\tau}$, the transition between these two regimes occurs at $\omega\tau \sim 1$ and is determined by the Lorentzian:

$$P(\omega\tau) = \frac{\tau^2}{1+(\omega\tau)^2}.$$

In Figure 5, for high ($\zeta = 3.5$, blue) and low ($\zeta = 0.001$, green) scaffold concentrations power spectra closely resemble the Lorentzian with the transition to $P(\omega\tau^{opt}) \sim \omega^{-2}$ behavior occurring at different frequencies. At low $\zeta = 0.001$ concentrations, the inverse time scale or corner frequency at which kinase activation decorrelates is determined by the diffusion limited rates of activation and deactivation of the final kinase C*. The corner frequency can be estimated from

$$S(t) \sim e^{-(t/\tau_c)},$$

where

$$(\tau_c)^{-1} = (k_+ + k_-) = O\,[D\,N_{tot}\,a].$$

$k_+$ and $k_-$ are diffusion limited rates of activation and deactivation and are given by a diffusion limited encounter rate that is on the order of $D\,N_{tot}\,a$ where $D$ is the diffusion constant used in the simulation, $N_{tot}$ is the number of proteins, and $a$ is the size of a protein taken to be the size of a lattice site. Substitution of the numbers used in the simulation (Table 1 and Methods) achieves a value for the relaxation time that is commensurate with the relaxation time for $\zeta = 0.001$ in Figure 2; i.e., $\tau_c \sim 10^5$ mcsteps.

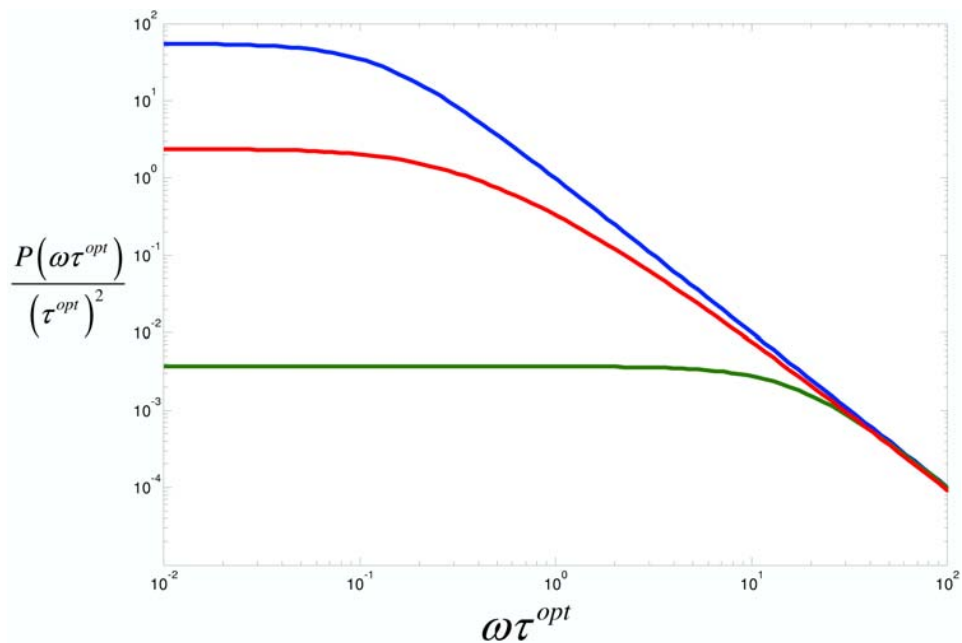

**Figure 5. Power spectra for kinase activation at high, low, and optimal scaffold concentrations.** Plots of $P(\omega) = |C(\omega)|^2$ ; $C(\omega) = \int e^{i\omega t} S(t) dt$ where $S(t) = e^{-(t/\tau)^\beta}$, are considered. Three cases are considered: low concentration ($\zeta = 0.005$, $\tau = 1.9 \times 10^5$ mcsteps, $\beta = 1.03$), high concentration ($\zeta = 3.5$, $\tau = 2.2 \times 10^7$ mcsteps, $\beta = 0.97$), and optimal concentration ($\zeta = 1.0$, $\tau = 3.0 \times 10^6$ mcsteps, $\beta = 0.60$). On the x-axis, frequency is reported in units that are scaled to the characteristic time for the $\zeta = 1.0$ case, $\tau^{opt} = 3.0 \times 10^6$ mcsteps. The y-axis contains values of $P(\omega\tau^{opt})/(\tau^{opt})^2$.
doi:10.1371/journal.pcbi.1000099.g005

At high $\zeta = 3.5$ concentrations, the corner frequency is determined by rates of formation and disassociation of an intact signaling complex. Furthermore, because of these many process that comprise the relaxation rate in this case, a numerical estimate of the corner frequency is difficult. In the case of the optimal ($\zeta = 1.0$, red) concentration, the transition from constant to $P(\omega \tau^{opt}) \sim \omega^{-2}$ behavior occurs smoothly over many decades from $\omega^{\tau opt} \sim 0.1$ to $\omega \tau^{opt} \sim 10.0$.

The plot in Figure 5 also resolves different frequency dependent processes occurring in signal transduction. At high frequencies or short times, $\omega \tau^{opt} < 10.0$, kinase activation is limited by the diffusive motion of the kinases in the cascade. At intermediate frequencies, $0.1 < \omega \tau^{opt} < 10.0$, activation is dominated by transitions between kinases assembled in competent, incompetent, and solution based kinases. For low frequencies $\omega \tau^{opt} < 0.1$ or long times, kinase activation decorrelates for each scaffold concentration.

## Relationship Between Signal Duration and Computed First Passage Time Statistics

To illustrate how computed values of $S(t)$ and the distribution of waiting times for kinase activation relate to conventional means of defining signal duration, we consider a differential equation for the time evolution of the activated form of the final kinase within the cascade. In this picture, species become activated at rates derived from the functional form that was fitted to the survival probabilities that were computed from the simulations. The waiting time or first-passage time distribution $f(t)$ is used as a forward rate and the activated final kinase then can be degraded with a kinetics of degradation characterized by a rate constant, $k_\phi$. A kinetic equation describing this process is written as:

$$\frac{dx}{dt} = \left\{ \frac{\beta(t/\tau)^\beta}{t} e^{-(t/\tau)^\beta} \right\} - k_\phi x.$$

$x$ is the number of active species, $\tau$ is the time constant of signal propagation, and $\beta$ is the stretching parameter that quantifies deviations away from single exponential behavior. In this picture, $x(t)$ represents the average response to a stimulus $f(t)$ that is distributed temporally according to $\frac{\beta(t/\tau)^\beta}{t} e^{-(t/\tau)^\beta}$ and subject to a first order decay with characteristic time $1/k_\phi$.

The equation for $x(t)$ can be solved and using the initial condition, $x(0) = 0$:

$$x(t) = \int_0^t e^{-k_\phi(t-t') - (t'/\tau)^\beta} \beta(t')^{-1} (t'/\tau)^\beta dt'.$$

$x(t)$ was integrated numerically and is shown for different values of $\beta$ in Figure 6A. As seen in Figure 5A, decreasing values of $\beta$ result in the trajectories having longer tails and thus an extended duration of signaling. Also, smaller values of $\beta$ result in the signal having a larger peak. This property directly follows from the decay of $S(t)$ that was shown in Figure 3A for different values of $\beta$. At early times, $S(t)$ decays more quickly when $\beta$ is smaller; as a consequence, more kinases are activated at these times, thus resulting in a larger peak.

This concept of signal duration can be made more precise by considering a threshold amount of signal, $T$, that is required for the pathway to be considered active. With a chosen value of $T$, the signal duration, $v$, is defined as the time it takes for the signal to decay to some threshold value, $T$. That is, the equation

$$T = \int_0^v e^{-k_\phi(v-t') - (t'/\tau)^\beta} \beta(t')^{-1} (t'/\tau)^\beta dt'.$$

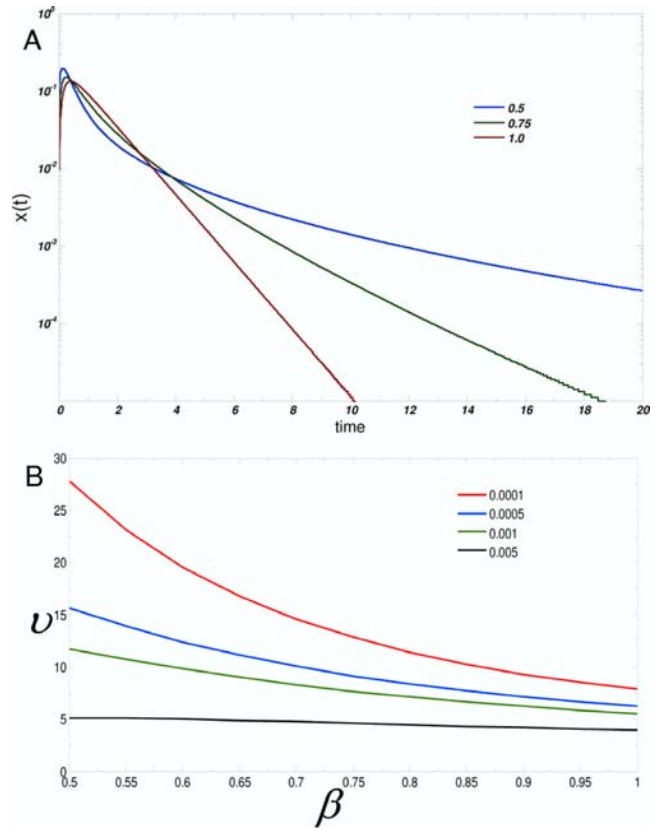

**Figure 6. Relation between first passage time statistics and signal duration.** (A) Trajectories of $x(t)$ on a semi-log plot. The abscissa represents time scaled by the characteristics time $\tau$. Trajectories for three values of $\beta$ are shown: $\beta = 0.06$ (blue), $\beta = 0.8$ (green), and $\beta = 1.0$ (red). $k_\phi = 5$ for each curve. Smaller values of $\beta$ result in larger values of $x(t)$ at longer times. (B) Values of signal duration as a function of $\beta$ for different choices of threshold, $T$ (defined in text); values of $T$ are provided in the legend. For small values of $T$, $\beta < 1$ (i.e., the presence of scaffolds) markedly increases signal duration.
doi:10.1371/journal.pcbi.1000099.g006

is satisfied. Figure 6B shows the signal duration, $v$, as a function $\beta$ for values of $\beta$ ranging from 0.5 to 1 for different values of $T$. For smaller values of $T$, $\beta < 1$ (i.e., scaffolds are present) results in a large increase in signal duration compared to the case in which $\beta = 1$. Therefore for a fixed value of $\tau$, the most broadly distributed signal leads to the longest signal duration.

## Discussion

We first showed that scaffold concentration is a key variable in regulating the speed of signal transduction. Moreover, we showed that the concentration of a scaffold protein can influence signaling dynamics by controlling the distribution of times over which kinases become active. This type of regulation may have many important consequences that are related to the influence of signal duration on cell decisions. Controlling the times over which kinases are activated may also be useful in directing a specific, robust response in a number of ways. Thus, the scaffold concentration itself provides another variable for maintaining signal specificity by controlling signal duration. This is consistent with data from genetic studies involving KSR1 [9,10], where the authors reported that the concentration of KSR1 can control a cell decision involving commitment to adipogenesis.

Our study focused solely on aspects of scaffold mediated regulation of signal transduction and we only considered the times at which kinases are active in the course of signal transduction. Many other factors also control signal duration. For example, our study does not consider the negative feedback loops that are often associated with the upregulation of phosphatases [18,32]) or the role of receptor downregulation in controlling signal duration. Also we did not explicitly consider the role of positive versus negative feedback loops in shaping signal duration which is undoubtedly important [30]. It was our focus to study how spatially localizing kinases on a scaffold protein influences signal duration. We aimed to untangle this effect of scaffold proteins from other essential features of kinase cascades such as allostery and feedback regulation. Also, other theoretical studies have investigated the first passage time statistics in signal transduction cascades and have found interesting dynamics that result from, in part, the sequential activation of multiple steps in a kinase cascade [35,36]. Our studies of signaling through scaffold proteins supplement these findings and, to our knowledge, provide the first study that shows how scaffolds affect the statistics of signal transduction.

Several predictions from our model of how scaffolds regulate signaling dynamics can be tested. Measurements that monitor the time course of signaling for different scaffold concentrations could potentially resolve the differences in signaling dynamics that are predicted. Also, single molecule or fluorescence correlation based spectroscopic methods [46–48] could potentially probe the statistics of signaling dynamics inherent in kinase cascades and study how such statistics are related to reliable cell decisions. Such techniques can monitor the propagation of a signal, at the level of an individual molecule and thus directly measure how kinase activation within a single cell is distributed over time.

## Methods

### Kinetic Monte Carlo Simulations

We simulate a model protein kinase cascade such as the mitogen-activated protein kinase (MAPK) cascade (Figure 1A) in the presence and absence of a scaffold with a kinetic Monte Carlo algorithm [49,50], which allows us to monitor the relevant stochastic dynamics. Since we are investigating phenomena that occurs on the time scales of signal transduction, we course-grain the system so that proteins are represented as discrete objects, occupying a site on a lattice of dimensions $100\times100\times100$ lattice spacings. Scaffold proteins are modeled as rigid, immobile objects containing three binding sites that are each specific for a particular kinase. When bound to a scaffold, kinases are tethered in nearest neighbor positions that are proximal to their downstream substrates. Allowing the scaffold and scaffold-bound species to move does not affect the qualitative results. Reflecting, no flux (i.e. Neumann) boundary conditions exist at each of the faces of the cubic lattice. The system is not periodically replicated since our simulation box is a size on the order a cell. Proteins can diffuse (i.e. translate on the lattice in random directions), bind and unbind, and undergo state transformations according to the prescribed reaction network involving a three staged cascade of activation and deactivation events (Figures 1A and 1B). Protein motion is subject to excluded volume (steric) constraints in that no two proteins can occupy the same site on the lattice. Chemical (state) transformations and binding events are modeled as thermally activated processes with energy barriers for activation, inactivation, binding and unbinding reactions. Parameters used are given in Table 1.

We simulate the dynamics with a fixed time step Monte Carlo algorithm. In a Monte Carlo step, n trials are attempted, where n is the number of proteins in the simulation. For a given trial, a protein is first chosen at random with uniform probability. A displacement move in a uniformly random direction is attempted with probability,

$$P(diffusion) = \frac{1}{2d} D^{eff} \min\{1, \exp(-E_\infty)\}$$

where d is the dimensionality of the simulation box, $D^{eff}$ is the probability of attempting a diffusion move and sets an overall time scale to diffuse the length of a lattice site. Excluded volume is accounted for by imposition of an infinite energy barrier, $E_\infty$, for hopping to sites containing other proteins; i.e.

$$E_\infty = \left\{ \begin{array}{ll} 0 & ;\ site\ is\ empty \\ \infty & ;\ site\ is\ occupied \end{array} \right\}.$$

Upon considering all possible nearest neighbor interactions, reaction moves, as determined by the network topology, are tried with probabililty,

$$P(reaction) = k^{eff} \min\{1, \exp(-E_{j,j'})\}$$

where $k^{eff}$ is the probability of attempting a reaction move; ($k^{eff}$ sets an overall reaction time scale), $E_{j,j'}$ is the energy barrier for the $j'\rightarrow j$ reaction scaled with respect to $k_b T$ (Boltzman's thermal energy). With this Monte Carlo move set, the simulations formally evolve the dynamics of the probability $P(\vec{r}|s^{i,j};t)$ that a chemical species $s^{i,j}$ of type $i$ and state $j$ at position $\vec{r}$ at time $t$ according to the Master equation

$$\frac{\partial}{\partial t} P(\vec{r}|s^{i,j};t) = \sum_{\vec{r}'} \omega^{i,j}(\vec{r}|\vec{r}') P(\vec{r}'|s^{i,j};t)$$

$$- \left[\sum_{\vec{r}'} \omega^{i,j}(\vec{r}'|\vec{r})\right] P(\vec{r}|s^{i,j};t)$$

$$+ \sum_{j'} \alpha(\vec{r};j|j') \Delta(\vec{r}-\vec{r}_k) P(\vec{r}|s^{i,j'};t)$$

$$- \left[\sum_{j'} \alpha(\vec{r};j'|j) \Delta(\vec{r}-\vec{r}_k)\right] P(\vec{r}|s^{i,j};t)$$

$$+ \sum_{\vec{r}'} \sum_{j''} \beta(\vec{r}|\vec{r}';j|j'') \Theta(\vec{r}';i'|i'';t) P(\vec{r}|s^{i,j''};t)$$

$$- \sum_{\vec{r}'} \sum_{j''} \beta(\vec{r}|\vec{r}';j''|j) \Theta(\vec{r}';i'|i'';t) P(\vec{r}|s^{i,j''};t)$$

where $\omega^{i,j}(\vec{r}|\vec{r}')$ is the transition probability per unit time for a displacement from $\vec{r}'$ to $\vec{r}$ of species $s^{i,j}$; $\omega^{i,j}(\vec{r}|\vec{r}') = \frac{D^{eff}}{2d} \min\{1, \exp(-E_\infty)\}$; $\alpha(\vec{r};j|j')$ is the per unit time transition probability at $\vec{r}$ for a species $s^{i,j'}$ to change to state $s^{i,j}$ (e.g. binding and unbinding reactions) and is $\alpha\left(\vec{r};j|j'\right)=k^{eff}\min\{1,\exp(-E_{j,j'})\}$, and $\Delta\left(\vec{r}-\vec{r}_k\right)$ imposes the constraint that binding and unbinding occurs only at specified binding sites on the scaffolds at positions $\vec{r}_k$ ($\Delta\left(\vec{r}-\vec{r}_k\right)$ is zero unless a scaffold is located at position $\vec{r}=\vec{r}_k$; and $\beta\left(\vec{r}|\vec{r}';j|j''\right)=k^{eff}\min\{1,\exp(-E_{j,j''})\}$ is the transition probability per unit time for a species at $\vec{r}'$ to facilitate (i.e. catalyze) the $j''\rightarrow j$ transformation at site $\vec{r}$; and $\Theta\left(\vec{r}';i'|i'';t\right)$ is zero unless the site at $\vec{r}'$ is occupied by the appropriate catalyst (i.e. $i'=i''$) in which case it is 1; each summation indicates a sum over nearest neighbors.

### Parameters Used

The parameters used in the simulation were first constrained to typical literature values. Energies of disassociation were taken to be $12k_bT$ corresponding of a disassociation constant $K_d$ of roughly 1 μM. 200 stimulatory molecules, S, 200 molecules of kinase A and a 1:1:5 ratio of A, B, and C kinases was used. If we assume a lattice spacing of 10nm, a typical diameter of a protein, the concentration of kinases in our simulation box is roughly 1 μM for kinase A and kinase B and ~5 μM for kinase C. In a physiological context, assuming the radius of the cell is about 10 μm, this approximately corresponds to ~$10^5$ molecules of kinases A and B and a copy number of ~$5\times10^5$ for kinase C in our simulation. 600 generic phosphatases are also present. These relative numbers are commensurate with reported kinase concentrations in Yeast and other systems [51,52]. Chemical kinetics were modeled in the simplest possible way by considering a single elementary reactive collision; i.e.,

$$A^* + B \rightarrow A^* + B^*$$

where the asterisk (*) denotes an active species. For the purposes of our simulations, saturation effects were ignored and the kinetics were taken to be in a linear regime. Such a model is reasonable when reactions are not limited by the availability of the enzyme. However, relaxing this assumption does not affect the qualitative behavior of our results provided that the times scales involved in the formation of an enzyme-substrate complex and subsequent catalysis do not compete with the diffusive processes in solution. If additional processes associated with enzyme catalysis dominate over diffusive motion of the proteins or binding and unbinding to and from the scaffold, then these process would be observed in the autocorrelation function and corresponding power spectrum. Given that catalysis would incorporate additional processes into the mechanism of kinase activation, such effects would serve to broaden the distribution at all scaffold concentrations as we have observed in our simulations (data not shown). We did not explore this scenario in its entirety since our aim was to solely investigate the effects of scaffolding a kinase cascade.

As discussed in a previous study [33], an important variable that determines the role of scaffolding a kinase cascade is the amount of time required ($\tau_{ec}$) for an active kinase to encounter its downstream target. For simple diffusion, in three dimensions, $\tau_{ec} \sim \dfrac{1}{DC^{2/3}}$ where $D$ is the diffusion constant and $C$ is a typical concentration of kinases. Experiments indicate that $\tau_{ec}$ is on the order of $10^{-4}$s–$10^0$ s [42]. Our studies focused on these experimentally relevant conditions.

Steady-state values are reported. The system is first placed in a random configuration and simulations are allowed to "equilibrate" by letting the dynamics evolve to a time much larger than the time it takes for a kinase to diffuse the length of the simulation box. Kinases that are inactive at time $t'$ are tagged and waiting times are observed at time $t+t'$ (i.e. statistics are collected for the times at which the kinases become activated), and $t'$ is chosen to be a time longer than the time required for equilibration of the Monte Carlo trajectory.

### Calculation of Statistical Quantities

Signaling dynamics can be defined microscopically as the distribution of times at which an individual kinase among of pool of available kinases becomes activated. Therefore, we quantify signaling dynamics by first considering the survival probability $S(t)$. $S(t)$ gives the probability that a particular kinase among the pool of signaling molecules has not been activated at time $t$ provided that it was inactive at time $t=0$. $S(t)$ is a two time point autocorrelation function:

$$S(t) = \langle \sigma(t)\sigma(0)\rangle,$$

where the brackets denote an ensemble average and $\sigma(t)$ is a binary variable indicating the state of a kinase; i.e.,

$$\sigma(t) = \left\{\begin{array}{l} 0 \; ; \; kinase\ is\ active \\ 1 \; ; \; kinase\ remains\ inactive \end{array}\right\}.$$

The survival probability is related to other dynamical properties; for instance, it can be related to a waiting time probability density function or first passage time distribution, $f(t)$, in the following way:

$$S(t) = \int_t^\infty f(t')dt' = 1 - \int_0^t f(t')dt' \quad and \quad f(t) = -\frac{d}{dt}S(t).$$

$S(t)$ is the complement of the cumulative probability distribution of the first passage time. $S(t)$ is computed from the simulations by integrating $f(t)$. Such a calculation is analogous to the data obtained from a single molecule experiment that measures the statistics of enzyme dynamics [46]. This distribution of waiting times underlies the intrinsic duration of signal propagation in a protein kinase cascade—the decay of such a quantity is a measure of how fast the signaling cascade responds to stimuli. Important to note is that this quantity gives information only on the timing of the signal and not on its final magnitude. We also consider the product of the survival probability with the probability that a kinase in the pool of signaling molecules is active in the steady state,

$$R(t) = f_A\langle \sigma(t)\sigma(0)\rangle,$$

where $f_A$ is the fraction of active kinases at steady state. When normalized, $R(t)$ is a measure of how the activity of the total pool of kinases is distributed over time, and can be thought of as an integrated flux of activated kinases. $-\dfrac{d}{dt}R(t)$ is seen as a reactive flux in provides a measure of the rate at which downstream kinases are being activated. One can imagine that both quantities could be biologically relevant. If conditions dictate that a biological response requires that a certain number of kinases remain active for extended amounts of time, $R(t)$ may be the more relevant quantity. On the other hand if the cellular decision requires a count of kinases that become active over a specified time window, then $S(t)$ could be the relevant quantity since it provides a measure of how the activation of individual kinases is distributed over time. Both quantities may be used to integrate signals in different contexts but since our study focuses on signaling dynamics we primarily focus on the survival probability and its related quantities.

Power spectra were computed numerically. Real and imaginary parts of the Fourier transform were obtained from numerical integration using the trapezoidal rule with a step size $\Delta t = 0.001$. $P(\omega)$ is calculated by squaring the real and imaginary parts of $X(\omega)$

$$P(\omega) = [\text{Re}X(\omega)]^2 + [\text{Im}X(\omega)]^2.$$

$P(\omega)$ was sampled at $N = 100$ logarithmically spaced (i.e., $\omega_{\max} = \omega_0 \ (10^{\delta(n-1)})$; $n \in [1,100]$ so that $\delta = \frac{1}{N-1}\log\left[\frac{\omega_{\max}}{\omega_0}\right]$) angular frequencies beginning at: $\omega_0 = \frac{2\pi}{T}$, where $T$ is the total length of the autocorrelation function.

## Author Contributions

Conceived and designed the experiments: JL. Performed the experiments: JL. Analyzed the data: JL. Contributed reagents/materials/analysis tools: JL. Wrote the paper: JL AC.